\documentclass[energies,article,accept,moreauthors,pdftex,12pt,a4paper]{mdpi} 

\lastpage{x}
\doinum{10.3390/------}
\pubvolume{xx}
\pubyear{2015}
\history{Received: xx / Accepted: xx / Published: xx}

\usepackage{graphicx}  
\usepackage{amsmath,amssymb,amsthm,mathrsfs,amsfonts,dsfont} 
\usepackage{accents}
\usepackage{algorithm,algorithmic}
\usepackage{booktabs}
\usepackage{txfonts}

\newcommand{\ubar}[1]{\underaccent{\bar}{#1}}

\Title{Learning Agent for a Heat-Pump Thermostat With a Set-Back Strategy Using Model-Free Reinforcement Learning}

\Author{Frederik Ruelens$^{1,3}$*, Sandro Iacovella$^{1,3}$, Bert J. Claessens$^{2,3}$, and Ronnie Belmans$^{1,3}$}

\address{%
$^{1}$ Faculty of Engineering, Department of Electrical Engineering, Division ELECTA, KU Leuven, Kasteelpark Arenberg 10, Box 2445, 3001 Leuven, Belgium;\\
$^{2}$ Energy Department of VITO, Flemish Institute for Technological Research, Boeretang 200, 2400 Mol, Belgium;\\
$^{3}$ EnergyVille, Thor park 8300, 3600 Genk, Belgium.\vspace{-0.3cm}}
\corres{ E-mail: frederik.ruelens@esat.kuleuven.be; Tel:+32 16 32 85 14.\vspace{-0.7cm}}
\abstract{The conventional control paradigm for a heat pump with a less efficient auxiliary heating element is to keep its temperature set point constant during the day.
This constant temperature set point ensures that the heat pump operates in its more efficient heat-pump mode and  minimizes the risk of activating the less efficient auxiliary heating element.
As an alternative to a constant set-point strategy, this paper proposes a learning agent for a thermostat with a set-back strategy.
This set-back strategy relaxes the set-point temperature during convenient moments, e.g. when the occupants are not at home.
Finding an optimal set-back strategy requires solving a sequential decision-making process under  uncertainty, which presents two challenges.
A first challenge is that for most residential buildings a description of the thermal characteristics of the building is unavailable and challenging to obtain.
A second challenge is  that the relevant information on the state, i.e. the building envelope, cannot be measured by the learning agent.
In order to overcome these two challenges, our paper proposes an auto-encoder coupled with a batch reinforcement learning technique.
The proposed approach is validated for two  building types with different thermal characteristics  for heating in the winter and cooling in the summer.
The simulation results indicate that the proposed learning agent can reduce the energy consumption  by $4{-}9\%$ during 100 winter days and by $9{-}11\%$ during 80 summer days compared to the conventional constant set-point strategy.
}
\vspace{-0.2cm}

\begin{document}


\renewcommand{\algorithmicrequire}{\textbf{Input:}}
\renewcommand{\algorithmicensure}{\textbf{Output:}}

\newcommand{\expected}{\mathop{\bf E\/}}
\newcommand{\Ex}{\mathop{\bf E\/}}

\newcommand{\uphk}{u_k^{\mathrm{ph}}}
\newcommand{\uph}{u^{\mathrm{ph}}}
\newcommand{\uphl}{u_l^{\mathrm{ph}}}
\newcommand{\xkt}{x_{k,\mathrm{t}}^{\mathrm{q}}}
\newcommand{\xkphys}{x_{k,\mathrm{ph}}}
\newcommand{\xkphysmin}{\underline{x}_{k,\mathrm{ph}}}
\newcommand{\xkphysmax}{\overline{x}_{k,\mathrm{ph}}}

\newcommand{\umax}{u_{\mathrm{max}}}
\newcommand{\tstart}{t_{\mathrm{s}}}
\newcommand{\tend}{t_{\mathrm{e}}}

\newcommand{\soc}{x_{\mathrm{soc}}}
\newcommand{\Tkin}{T_{k,\mathrm{in}}}
\newcommand{\Tkout}{T_{k,\mathrm{out}}}
\newcommand{\batchUphys}{ \mathcal{F} = \{(x_{l},u_{l},x_{l}',\uphl)\}_{l=1}^{\#\mathcal{F}}}
\newcommand{\batchCost}{ \mathcal{F} = \{(x_{l},u_{l},x_{l}',c_{l})\}_{l=1}^{\#\mathcal{F}}}
\newcommand{\batchUphysbold}{ \mathcal{F} = \{(\mathbf{x}_{l},u_{l},\mathbf{x}_{l}',\uphl)\}_{l=1}^{\#\mathcal{F}}}

\newcommand{\batchUphysboldR}{ \mathcal{F}_{ \mathcal{R}} = \{(\mathbf{x}_{l},u_{l},\mathbf{x}_{l}',\uphl)\}_{l=1}^{\#\mathcal{F}}}

\newcommand{\batchR}{ \mathcal{F}_{ \mathcal{R}} }
\newcommand{\batch}{ \mathcal{F}_{ \mathcal{R}} }

\newcommand{\tuple}{(x_{l},u_{l},x_{l}',\uphl)}

\newcommand{\forecastexo}{\hat{X}_{\mathrm{ex}}}
\newcommand{\forecastexoPhys}{\hat{X}_{\mathrm{ex}}^{\mathrm{ph}}}
\newcommand{\forecastexoCost}{\hat{X}_{\mathrm{ex}}^{\mathrm{c}}}

\newcommand{\probwk}{p_{\mathcal{W}}(\cdot|x_{k})}
\newcommand{\probw}{p_{\mathcal{W}}(\cdot|x)}
\newcommand{\Tin}{T_{\mathrm{in}}}
\newcommand{\Tint}{T_{\mathrm{in},k}}
\newcommand{\Tout}{T_{\mathrm{out}}}
\newcommand{\Toutt}{T_{\mathrm{out},k}}
\newcommand{\Hm}{H_{\mathrm{m}}}
\newcommand{\Cm}{C_{\mathrm{m}}}
\newcommand{\Ca}{C_{\mathrm{a}}}
\newcommand{\Ua}{U_{\mathrm{a}}}
\newcommand{\Tm}{T_{\mathrm{m}}}
\newcommand{\Qi}{Q_{\mathrm{i}}}
\newcommand{\Qg}{Q_{\mathrm{g}}}
\newcommand{\Qm}{Q_{\mathrm{m}}}
\newcommand{\Tindot}{\dot{T}_{\mathrm{i}}}
\newcommand{\Tmdot}{\dot{T}_{\mathrm{m}}}
\newcommand{\Qs}{Q_{\mathrm{s}}}
\newcommand{\Pheat}{P_{\mathrm{h}}}
\newcommand{\Pcool}{P_{\mathrm{c}}}
\newcommand{\Pheatmax}{P_{\mathrm{h}}^{\mathrm{max}}}
\newcommand{\Pcoolmax}{P_{\mathrm{c}}^{\mathrm{max}}}
\newcommand{\minimum}{\mathrm{max}}
\newcommand{\Uphys}{U^{\mathrm{ph}}}

\newcommand{\Qhp}{Q_{\mathrm{h}}}
\newcommand{\Paux}{P_{\mathrm{a}}}
\newcommand{\Tsett}{T_{\mathrm{s},k}}
\newcommand{\Tset}{\ubar{T}_{\mathrm{s}}}
\newcommand{\Tsetmax}{\bar{T_{\mathrm{s}}}}
\newcommand{\Taux}{T_{\mathrm{a}}}
\newcommand{\Tsetback}{T_{\mathrm{sb}}}
\newcommand{\Tbuffer}{T_{\mathrm{b}}}
\newcommand{\Tbufferaux}{T_{\mathrm{b}}^{\mathrm{a}}}

\newcommand{\nnim}{n_{\mathrm{min}}}
\newcommand{\Solart}{S_{k}}
\newcommand{\nsteps}{n_{\mathrm{s}}}
\newcommand{\Tintrace}{T_{\mathrm{in},k-1},...,T_{\mathrm{in},k-n}}
\newcommand{\utrace}{u_{k-1}^{\mathrm{ph}},...,u_{k-n}^{\mathrm{ph}}}
\newcommand{\zenc}{\mathbf{z}_{\mathrm{e}}}
\newcommand{\zenck}{\mathbf{z}_{\mathrm{e},k}}

\newcommand{\Tsetvalue}{20.5^{\circ}\mathrm{C}}
\newcommand{\Tbuffervalue}{0.5^{\circ}\mathrm{C}}
\newcommand{\Tbufferauxvalue}{1.5^{\circ}\mathrm{C}}
\newcommand{\days}{8}
\newcommand{\atWork}{7}
\newcommand{\atHome}{17}
\newcommand{\result}{7\%}

\newcommand{\resultwinter}{4\text{-}9\%}
\newcommand{\resultsummer}{7\text{-}11\%}
\newcommand{\resultwinterkwh}{500\mathrm{kWh}}
\newcommand{\resultsummerkwh}{900\mathrm{kWh}}

\newcommand{\Xtime}{X_{\mathrm{t}}}
\newcommand{\Xphys}{X_{\mathrm{ph}}}
\newcommand{\Xexo}{X_{\mathrm{ex}}}

\section{Introduction}

Residential and commercial buildings use about $20{-}40\%$ of the global energy consumption~\cite{IEA_online}.
Half of this energy is consumed by Heating, Ventilation and Air Conditioning  (HVAC) systems. 
About two-thirds of these HVAC systems use fossil fuel  sources, such as oil, coal and natural gas.
Replacing this large share of  fossil fueled HVAC systems with more energy efficient heat pumps can play an important role in reducing greenhouse gasses~\cite{Moretti,luickx2008influence,forsen2005heat}.
For instance in~\cite{bayer2012greenhouse}, Bayer \textit{et al.} report  that replacing fossil fuel based HVAC systems with electric heat pumps can help reduce greenhouse gasses in space heating by $30{-}80\%$ in different European countries.
The cardinal factors that influence this reduction are the substituted fuel type, the energy efficiency of the heat pump and the electricity generation mix of the country.

This paper focuses on  residential heat pumps  equipped with an  auxiliary heating element.
This heating element can be a less efficient electric furnace or a  gas- or oil-fired furnace.
In its regular operation, a heat pump runs in its more energy efficient heat-pump mode, however, when the temperature drops too low, both the heat pump  and the auxiliary heating element are activated. 
Since most heat pumps are equipped with an electric auxiliary heating element, which can be four times less efficient, the U.S. Department of Energy recommends to operate the thermostat with a constant target temperature during the day, even when the inhabitants are not at home~\cite{DoE}.

As an alternative to the constant temperature set-point strategy, this paper presents a set-back method, in which the temperature set point is relaxed during convenient times, for example, during the night or when the inhabitants are not at home.
Such a set-back method can reduce the energy consumption compared to the constant set-point strategy under the condition that it can avoid the auxiliary heating to activate~\cite{Savings}.

The remainder of this paper is organized as follows. 
Section~\ref{litreview} gives on overview of existing literature on heat-pump thermostats and their application to demand response.
Section~\ref{ps} addresses the challenges of developing a successful set-back strategy to reduce the energy consumption of a heat pump.
Section~\ref{mdp} formulates the sequential decision-making problem of a thermostat agent as a  stochastic Markov decision process.
Section~\ref{fqi} proposes an approach based on an auto-encoder and fitted Q-iteration.
The simulation results are given in Section~\ref{sr}, and, finally,  Section ~\ref{cfw} summarizes the general conclusion of this work.

\section{Literature review}
\label{litreview}

Driven by the potential of heat pumps to reduce greenhouse gasses, heat-pump thermostats have attracted the attention from  researchers~\cite{urieli2013learning,rogers2011adaptive} and commercial companies~\cite{Nest,Honneywell,buildingiq,NeuroBat,ANNA}. 
A popular control paradigm in the literature on optimal control of heat-pump thermostats is a model-based approach.
Within this paradigm, a first type of model-based controllers uses a model predictive control approach~\cite{Moretti,Treado}.
At each decision step, the controller defines a control action by solving a fixed-horizon optimization problem, starting from the current time step and using a calibrated model of its environment.
For example, the authors of~\cite{rogers2011adaptive},  use a mixed-integer quadratic programming solution to minimize the electricity cost and carbon output of a home heating system.
However, the performance of these model-based approaches depends on the quality of the model and the availability of expert knowledge.
Model-based approaches can achieve very good results within a reasonable learning period, but typically they have to be tailored for their application and they have difficulties with stochastic environments~\cite{challengesMPC}.
A second type of model-based controllers formulates the control problem as a Markov decision problem and solves the corresponding problem using techniques from approximate dynamic programming~\cite{powell2007approximate,bertsekas1995dynamic}.
For example in~\cite{urieli2013learning}, Urieli $\textit{et al.}$ use a linear regression model to fit the model of the building and then apply a tree-search algorithm for finding an intelligent set-back strategy for a heat-pump thermostat.
Alternatively in~\cite{morel2001neurobat}, Morel $\textit{et al.}$ propose an adaptive building controller that makes use of artificial neural networks and dynamic programming.
Similarly, the authors of~\cite{collotta2014dynamic} propose a combined neuro-fuzzy model for dynamic and automatic regulation of indoor temperature.
They use an artificial neural network to forecast the  indoor temperature, which is then used as the input of a fuzzy logic control unit in order to manage energy consumption of the HVAC system.
In addition, the authors of~\cite{moon2013determining} report that an artificial neural network based model can adapt to changing building background conditions, such as the building configuration, without the need for additional intervention by an expert.

An alternative control paradigm makes use of model-free reinforcement learning  techniques in order to avoid the system identification step of model-based controllers.
For example, the authors of~\cite{henze2003evaluation}, propose a Q-learning approach to minimize the electricity cost of a thermal storage. 
In~\cite{wen2014optimal}, Zheng \textit{et al.} show how a Q-learning approach can be used for residential and commercial buildings by  decomposing it over different device clusters.
However, a main drawback of classic reinforcement learning algorithms, such as Q-learning and SARSA, is that they discard observations  after each interaction with their environment.
In contrast, batch reinforcement learning techniques do not require many interactions until convergence to obtain reasonable policies~\cite{ernst2005tree,ernst2009reinforcement,lange2012batch}, since they store and reuse past observations. 
As a result, they have a shorter learning period which makes them an attractive technique for real-world applications, such as a heat-pump thermostat.
In both~\cite{claessens2013peak} and~\cite{RuelensBRLCluster}, the authors use a bath reinforcement learning  technique, fitted Q-iteration, in combination with a market-based multi-agent system, in order to control a cluster of flexible devices, such as electric vehicles and electric water heaters.

This work contributes to the application of batch reinforcement learning to the problem of finding a successful set-back strategy for a heat-pump thermostat.
This problem was previously addressed by the work of Urieli $\textit{et al.}$ in~\cite{urieli2013learning}.  
The main difference with their work is that our work proposes a model-free approach that can intrinsically capture the stochastic nature of the problem. 
The authors build on the existing literature on batch reinforcement learning, in particular fitted Q-iteration~\cite{ernst2005tree}, and auto-encoders~\cite{lange2010deep}.

\section{Problem Statement}

\label{ps}
\begin{figure}[t]
   \begin{center} 
			\includegraphics[width=0.60\columnwidth]{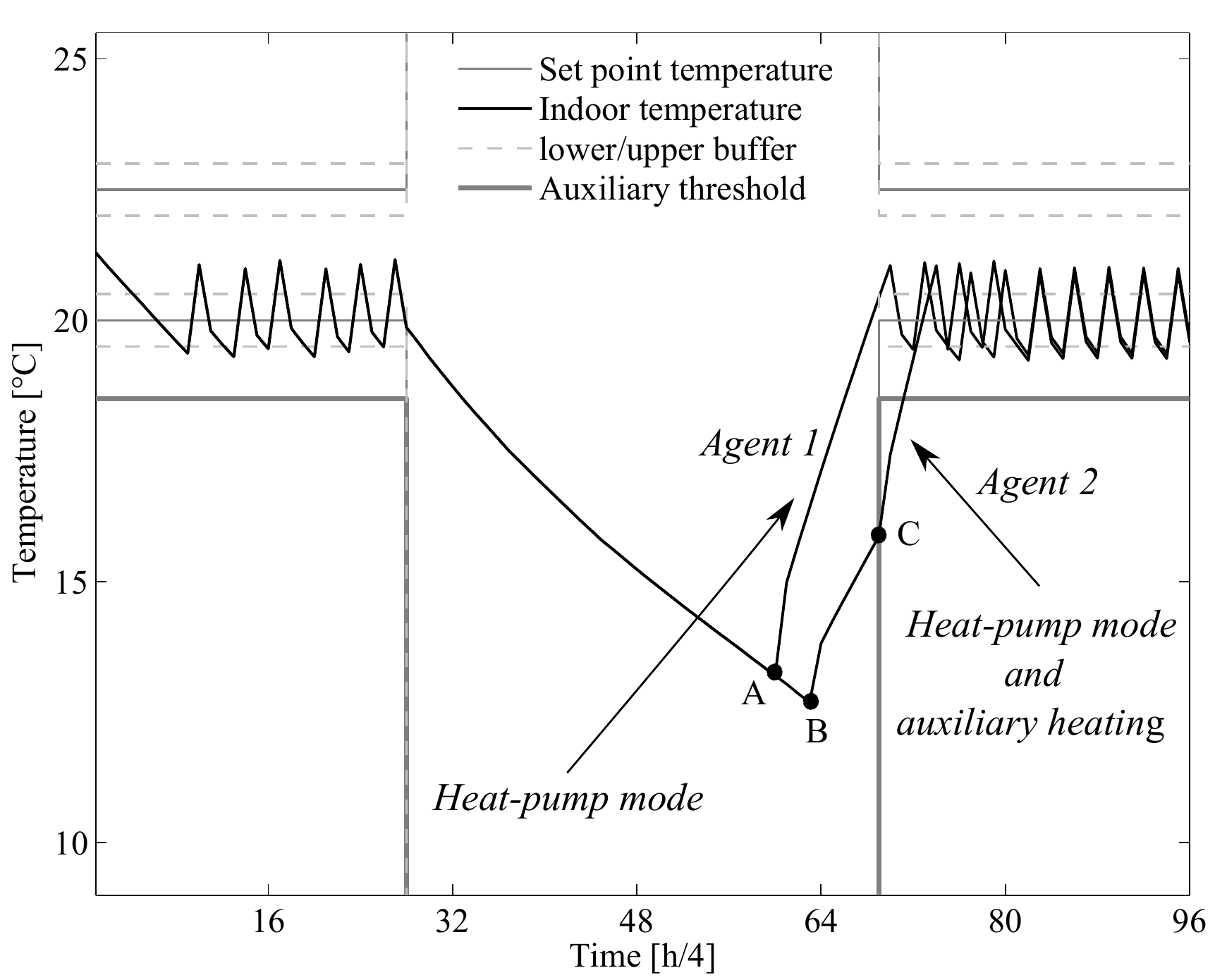}
\caption{Indoor temperature of two thermostat agents with a set-back strategy (7-17h). Agent one operates in normal heat-pump mode and agent two activates the less efficient auxiliary heating.}
\label{fig.AUXvsHP}
 \end{center}
\end{figure}

The main objective of this paper is to develop a model-free learning agent for a heat pump with an auxiliary heating element in order to overcome the following two challenges.
The first challenge is that the auxiliary heating element activates when the indoor temperature reaches a predefined temperature threshold. 
The operation of the thermostat is given by Algorithm~\ref{default} and  can be found in Appendix A.
More information in the  temperature settings of the thermostat can be found in Table~\ref{hp_paramters} of Appendix A.
In order to illustrate the activation of the auxiliary heating element, two thermostat agents are depicted in Figure~\ref{fig.AUXvsHP}.
Our set-back strategy  relaxes the indoor temperature during working hours, i.e. 7-17h (Figure~\ref{fig.AUXvsHP}).
It can be seen that the first agent correctly anticipates the comfort bounds at $\atHome$h and begins to heat the building in normal heat-pump operating mode (point A). 
The second agent postpones heating until point B and triggers the electric auxiliary heating to switch on in point C.
As a result of the activation of the less efficient auxiliary heating element, the second agent consumes more energy than the recommended constant temperature set-point strategy.

\begin{figure}[t]
   \begin{center} 
			\includegraphics[width=0.65\columnwidth]{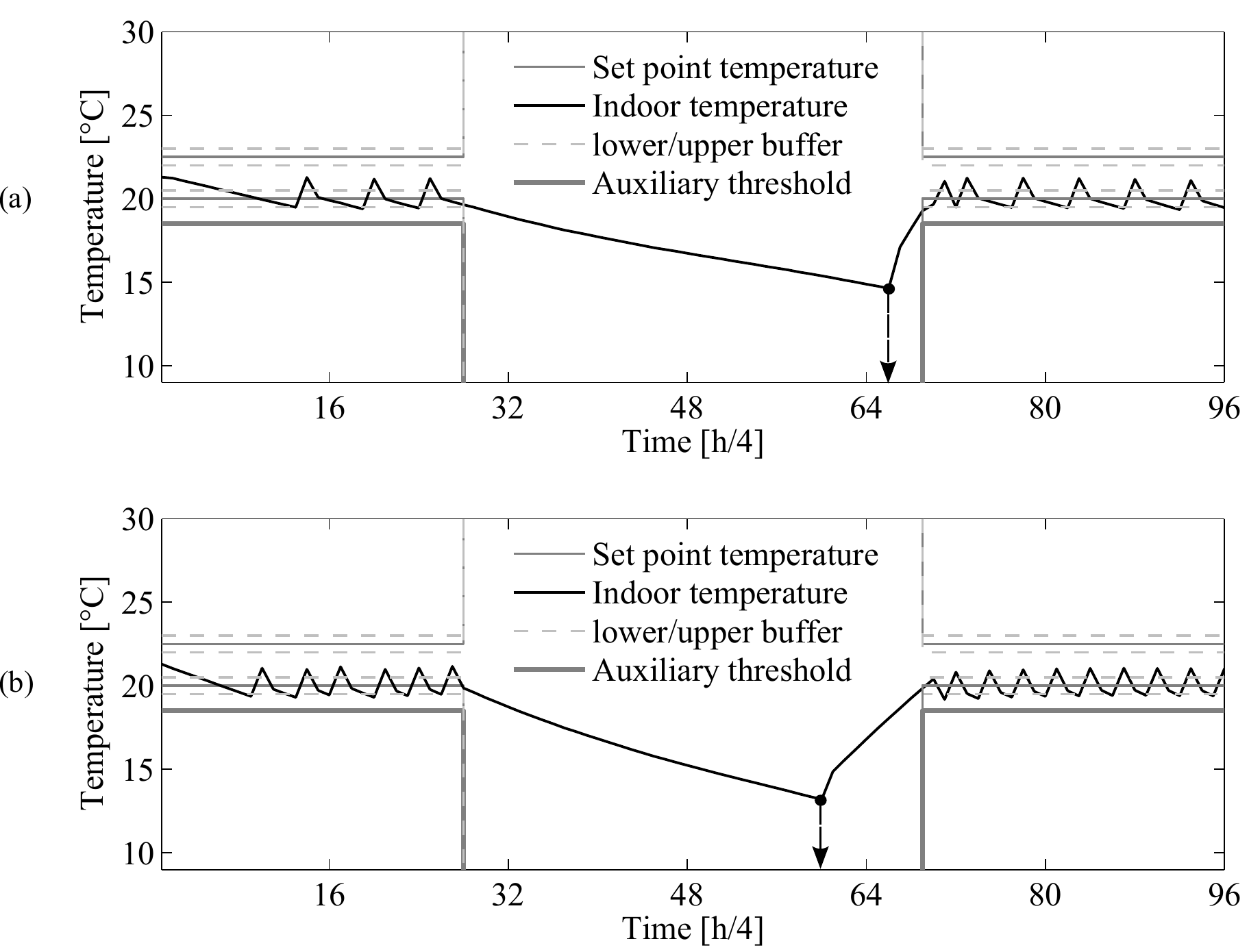}
\caption{Indoor temperature of two buildings with different thermal insulation. (a): building with high thermal insulation level. (b): building with low thermal insulation level.}
\label{insulation}
 \end{center}
\end{figure}

A second important challenge when developing an intelligent set-back strategy is that the moment of activating the heat pump does not only depend on the weather conditions, but also on the thermal characteristics of the building.
This challenge is illustrated by a second example, where a successful set-back strategy is depicted for two building types.
Both building types have identical outside temperatures and inner disturbances.
Figure~\ref{insulation}a. depicts the indoor temperature of a building with a  high insulation level, whereas Figure~\ref{insulation}b.  depicts the indoor temperature of a building with a low insulation level.
It can be seen that the thermal characteristics of the building can  have a significant impact on the operation of the thermostat agent.
For instance, the set-back thermostat in Figure~\ref{insulation}a. can postpone its heating action until quarter 68, while  the set-back thermostat in Figure~\ref{insulation}b. needs to start heating around quarter 60 in order to avoid the activation of the auxiliary heating.

\section{Markov Decision Process}
\label{mdp}

Motivated by the challenges presented in Section~\ref{ps} and driven by recent advances in reinforcement learning~\cite{ernst2005tree,deepmind,riedmiller2009reinforcement}, our paper introduces a model-free learning agent. 
In order to use reinforcement learning techniques,  the sequential decision-making problem of a heat-pump thermostat with set-back strategy is formulated as a stochastic Markov decision process~\cite{sutton1998reinforcement,bertsekas1995dynamic}.

At every decision step $k$ the thermostat agent chooses a control action $u_{k} \in U \subset \varmathbb{R} $ and the state of its environment $x_{k} \in X \subset \varmathbb{R}^{d}$ evolves according to the transition function $f$:
\begin{equation}
x_{k+1} = f(x_{k},u_{k},w_{k}) ~~ \forall k \in \{1,...,T-1\},
\label{eq.f}
\end{equation} with $w_{k}$  a realization of a random process drawn from a conditional probability  distribution $\probwk$.
After a transition to the next state $x_{k+1}$, the agent receives an immediate cost $c_{k}$ provided by:
\begin{equation}
c_{k}=\rho(x_{k},u_{k},w_{k}) ~~ \forall k \in \{1,...,T\},
\label{eq.reward}
\end{equation}
where $\rho$ is the cost function.
The goal of the thermostat agent is to find a  control policy ${h^{*}:X\rightarrow U}$ that minimizes the expected $T$-stage return for any state in the state space.
The expected $T$-stage return $J^{h^{*}}_{T}$ starting from $x_{1}$ and following $h^{*}$ is defined as follows:
\begin{equation}
J^{h^{*}}_{T}(x_{1}) = \underset{w_{k}\sim\probwk}{\Ex} \left[  \sum_{k=1}^{T}{\rho(x_{k},h^{*}(x_{k}),w_{k})}\right],
\end{equation}
where ${\Ex}$ denotes the expectation operator.
A more convenient way to  characterize the policy $h^{*}$ is  by using  a state-action value function or Q-function:
\begin{equation}
Q^{h^{*}}(x,u) = \underset{w\sim\probw}{\Ex} \left[\rho(x,u,w) +J^{h^{*}}_{T}(f(x,h(x),w)) \right].
\label{Qfunction}
\end{equation}
The Q-value is the cumulative return starting from  state $x$, taking action $u$, and following $h^{*}$ thereafter.
Starting from a Q-function for every state-action pair, the policy is calculated as follows:
\begin{equation}
h^{*}(x)  \in \underset{u \in U}{\text{arg min~}} Q^{h^{*}}(x,u),
\label{Qpolicy}
\end{equation}
where $h^{*}$ satisfies the Bellman optimality equation~\cite{BellmanDP}:
\begin{equation}
J^{h^{*}}(x) =  \underset{u \in U}{\min} \underset{w\sim\probw}{\Ex} \left[ \rho(x,h(x),w) + J^{h^{*}}_{T}(f(x,h(x),w)) \right].
\label{belli}
\end{equation}

The central idea behind batch reinforcement learning is to estimate the state-action value function $Q^{h^{*}}$ based on a set of past observations (or batch)  of the state, control action and reward.
Note that this approach does not require a model of the environment $f$ or the disturbances $w$. As a result no system identification step is needed.
The following five paragraphs give a tailored definition of the state, action, cost function and transition function of a heat-pump thermostat agent.


\subsection{Observable State} 

At each time step $k$, the thermostat agent can measure the following state information:
\begin{equation}
\mathbf{x}_{k} = (d,t,\Tint,\Toutt,\Solart),
\end{equation}
where $d \in \{1,\ldots, 7\}$ represents the current day in the week and $t \in \{1,\ldots, 96\}$  the current quarter of the hour. 
The observable state information related to the physical state of the building is given by  a measurement of the indoor temperature $\Tint$.
The observable exogenous state information is defined by $\Toutt$ and $\Solart$,which are the outdoor temperature  and  solar irradiance  at time step $k$.
Note that by including the measurements of $\Toutt$ and $\Solart$ at time step $k$ our approach captures a first-order correlation of these stochastic variables.

\subsection{Thermostat Function}
\label{subsection.backup_controller}

In order to guarantee the comfort of the end-user the heat pump is equipped with a thermostat mechanism (Algorithm~\ref{default}).
The thermostat logic  maps the requested control action $u_{k}$ taken in  state $\mathbf{x}_{k}$ to a physical control action $\uphk$: 
\begin{equation}
\uphk= T(\mathbf{x}_{k},u_{k}).
\end{equation}
 As such, the thermostat function $T$ maps the requested control action to a physical quantity, which is required to calculate the cost value.

\subsection{Augmented State}

As previously stated, this paper assumes that the temperature of the building envelope cannot be measured.
It is important to realize that the temperature of the building envelope contains essential information to accurately capture the transient response of the indoor air temperature.
Moreover, the temperature of the building envelope represents information on the amount of  thermal energy stored in the thermal mass of the building. 
A possible strategy is to represent the temperature of the building envelope by a handcrafted feature based on expert knowledge, which can be difficult to obtain for residential buildings.
However, a more generic strategy is to include past observations of the state and action in the state variable~\cite{deepmind,bertsekas1996neuro}:
\begin{equation}
\mathbf{x}_{\mathrm{aug},k} = (d,t,\Tint,\mathbf{z}_{k},\Toutt,\Solart),
\label{xaug}
\end{equation}
with 
\begin{equation}
\mathbf{z}_{k} = (\Tintrace,\utrace),
\end{equation} 
where $n$ denotes the number of past observations of the indoor temperatures and  physical actions.
Note that the physical control actions have been included in the state, since they give an indication of the amount of energy added to the system.
In the next section,  a feature extraction technique is proposed to mitigate the ``curse of dimensionality''~\cite{BellmanDP} and find a compact representation of the augmented state vector. 

\subsection{Transition Function} 

A detailed description of the transition function $f$ that models the temperature dynamics of the building is given in Appendix B.
This paper proposes a model-free approach and makes no assumption of the model type or its parameters.

\subsection{Cost Function} 

The cost function ${\rho:X \times U\rightarrow \varmathbb{R}}$,  associated with a single transition, is given by:
\begin{equation}
c_{k} = \uphk\Delta t+ \alpha_k,
\end{equation}
where the parameter $\alpha_k$ represents a penalty for violating the  comfort constraints and $\Delta t$ represent the time interval of one control period.
When the indoor air temperature $\Tin$ is lower than $\Tset$ or higher than $\Tsetmax$, $\alpha_k$  is set to $10^{5}$ and otherwise 0.

\section{Model-Free Batch Reinforcement Learning Approach}
\label{fqi}

\newcommand{\batchUphysboldaug}{ \mathcal{F} = \{(\mathbf{x}_{\mathrm{aug},l},u_{l},\mathbf{x}_{\mathrm{aug},l}',\uphl)\}_{l=1}^{\#\mathcal{F}}}

\begin{figure}
\centering
\includegraphics[width=0.75\columnwidth]{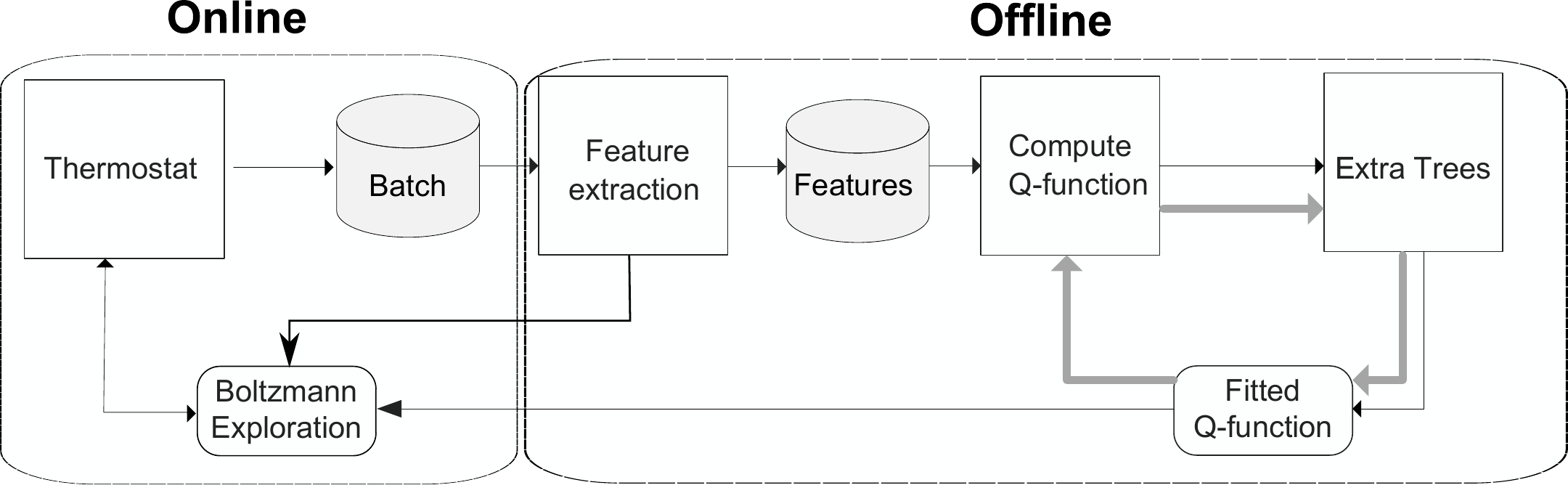}
\caption{Online and offline loop of our proposed approach, which consists of a feature extraction (offline), fitted Q-iteration (offline) and Boltzmann exploration (online).}
\label{fig.fqiScheme}
\end{figure}

Given full knowledge of the transition function, it can be possible to find an optimal policy by solving the Bellman equation~(\ref{belli}) for every state-action pair using techniques from approximate dynamic programming~\cite{powell2007approximate,BellmanDP}.
This paper, however, applies a model-free batch reinforcement learning technique, where the sole information available to solve the problem is the one obtained from daily observations of the following one-step transitions:
\begin{equation}
\batchUphysboldaug,
\end{equation}
where each tuple is made up of the augmented state $\mathbf{x}_{\mathrm{aug},l}$, the control action $u_{l}$, physical control action $\uphl$ and its successor state  $\mathbf{x}_{\mathrm{aug},l}'$.
Figure~\ref{fig.fqiScheme} outlines the building blocks of the model-free batch reinforcement learning method, which consists of two interconnected loops. 

\subsection{Offline Loop}

The offline loop contains a feature extraction technique and a batch reinforcement learning method.
\subsubsection{Feature Extraction}
This paper proposes a feature extraction technique to find a low dimensional representation of the augmented state
$\mathbf{x}_{\mathrm{aug},k} = (d, t,\Tint,\mathbf{z}_{k},\Toutt,\Solart)$, by reducing the dimensionality of the state information corresponding to past observations:
\begin{equation}
\zenck= \mathrm{\Phi}(\mathbf{z}_{k},W),
\end{equation}
where ${\mathrm{\Phi} :Z \rightarrow Z_{\mathrm{e}}}$ is  a feature extraction function  that maps $\mathbf{z} \in  Z \subset \varmathbb{R}^{p}$ to the encoded state $\zenc \in  Z_{\mathrm{e}} \subset \varmathbb{R}^{q}$, with  $q < p$ and where $W$ contains the parameters corresponding to $ \mathrm{\Phi}$.

This work  introduces a  feature extraction technique based on an auto-encoder.
An auto-encoder or auto-associative neural network is an artificial neural network, with the same number of input as output neurons and a smaller number of hidden feature neurons.
These hidden feature neurons function as a bottleneck and can be seen as a reduced representation of $\mathbf{z}_{k}$.
During training of the auto-encoder the output data is set to be equal to the input data.
The weights of the network are then trained to minimize the squared  error between the inputs and its reconstruction.
Different training methods to find $W$ can be found in the literature~\cite{hinton2006reducing,riedmiller1993direct,hestenes1952methods}.
However, comparing the performance of these training methods is out of the scope of this paper.
This work uses an hierarchical training strategy that uses a conjugate gradient descent method~\cite{Scholz}.

\newcommand{\batchUphysboldaugred}{ \mathcal{F}_{\mathcal{R}} = \{(\mathbf{\hat{x}}_{\mathrm{aug},l},u_{l},\mathbf{\hat{x}}_{\mathrm{aug},l+1},\uphl)\}_{l=1}^{\#\mathcal{F}}}

\newcommand{\Tinl}{T_{\mathrm{in},l}}
\newcommand{\Toul}{T_{\mathrm{out},l}}
\newcommand{\Solartl}{S_{l}}

The next paragraph explains how a popular batch reinforcement learning technique, i.e. fitted Q-iteration, can be used, given:
\begin{equation}
\batchUphysboldaugred,
\end{equation}
with
\begin{equation}
\mathbf{\hat{x}}_{\mathrm{aug},l} = (d, t,\Tinl,\mathrm{\Phi}(\mathbf{z}_{l},W),\Toul,\Solartl).
\label{augmentedState}
\end{equation}
where $\mathbf{\hat{x}}_{\mathrm{aug},l}$  denotes the reduced augmented state, and $\mathrm{\Phi}(\mathbf{z}_{l},W)$ denotes the encoded state information of the past observations $\mathbf{z}_{l}$.

\subsubsection{Fitted Q-iteration}
Although other batch reinforcement learning techniques can be used, this work contributes to the application of fitted Q-iteration~\cite{ernst2005tree}.
Fitted Q-iteration makes efficient use of gathered data and can be combined with different regression methods.
In contrast to standard Q-learning~\cite{sutton1998reinforcement}, fitted Q-iteration  computes the Q-function offline and makes use of the whole batch.
An overview of the fitted Q-iteration algorithm is given in Algorithm~\ref{FQI}.
The algorithm iteratively builds a training set with all  state-action pairs in $\mathcal{F}_{\mathcal{R}}$ as the input.
The target values consist of the corresponding cost values  and the optimal Q-value, based on the approximation of the previous iteration, for the next state.
This works uses an  extremely randomized trees ensemble method~\cite{geurts2006extremely} to find an approximation $\widehat{Q}$ of the Q-function.  
The ensemble was set to 60 trees and a minimum of 3 samples for splitting a node.
The number of samples selected at each node was set to the input dimension of the input space.
More information on the regression method can be found in~\cite{geurts2006extremely}. 

\subsection{Online loop}

A Boltzmann exploration strategy~\cite{kaelbling1996reinforcement} is used at each decision step to find  the  probability of selecting an action:
\begin{equation}
P\left(u|\mathbf{\hat{x}}_{\mathrm{aug},k}\right) = \frac{e^{Q^{*}\left(\mathbf{\hat{x}}_{\mathrm{aug},k},u\right)/\tau_{d}}}{\Sigma_{u'\in U} e^{Q^{*}\left(\mathbf{\hat{x}}_{\mathrm{aug},k},u'\right)/\tau_{d}}},
\end{equation}
where the parameter $\tau_{d}$ controls the amount of exploration and $Q^{*}$ is the Q-function obtained with Algorithm~\ref{FQI}.
The parameter $\tau_{d}$ is decreased during the simulation following an harmonic sequence~\cite{powell2007approximate}:
\begin{equation}
\tau_{d} = \frac{1}{(d)^{n}},
\end{equation}
where $d$ denotes the current day and $n$ is set to 0,7.
Note that, if $\tau_{d} = 0$, than the policy become greedy and the best action is chosen.

\begin{algorithm}[t]
\caption{Fitted Q-iteration~\cite{ernst2005tree}}
\label{FQI}
\begin{algorithmic}[1] 
\algsetup{linenosize=\tiny}
\renewcommand{\algorithmicrequire}{\textbf{Input:}}
\REQUIRE $ \batchUphysboldaugred$\\
\STATE Initialize $\widehat{Q}_{0}$  to zero
\FOR {$N = 1,\ldots,T$}
\FOR {$l = 1,\ldots,\#\mathcal{F}$}
\STATE $~~c_{l} = \rho(\mathbf{\hat{x}}_{\mathrm{aug},l},\uphl) $
\STATE $~~Q_{N,l}\leftarrow c_{l}+\underset{u \in U}{\text{min~}}\widehat{Q}_{N-1,l}(\mathbf{\hat{x}}_{\mathrm{aug},l+1},u) $
\ENDFOR
\STATE use regression to obtain $\widehat{Q}_{N}$ from $\left\{\left((\mathbf{\hat{x}}_{\mathrm{aug},l},u_{l}),Q_{N,l}\right),l =1,\ldots,\#\mathcal{F}\right\}$
\ENDFOR
\ENSURE $Q^{*}=\widehat{Q}_{N}$
\end{algorithmic}
\end{algorithm}

\newcommand{\energySavingsLow}{6\text{-}8\%}
\newcommand{\energySavingsHigh}{7\text{-}9\%}
\newcommand{\energySavings}{450}

\section{Simulation Results}
\label{sr}

This section compares the performance of our learning agent to a default constant  set-point strategy and a prescient set-back strategy.

\subsection{Simulation Setup}

The simulations use a second-order equivalent thermal parameter model to calculate the indoor air temperature and the envelope temperature of the building~\cite{chassin2008gridlab}.
The parameters correspond to a building with a floor area of  200 $\mathrm{m}^2$ and a window to floor ratio of $30\%$.
The model equations and parameters are presented in  Appendix B.
The building is equipped with a heat pump to satisfy the heating or cooling demand of the inhabitants.
The heat pump can change its power set point every 15 minutes with 10 discrete heating or cooling actions. 
This paper considers two comfort settings, i.e. the default strategy and the set-back strategy.
The default strategy has a constant temperature set point of $\Tsetvalue$ during the entire day. 
In contrast, the set-back strategy relaxes the set-point  temperature from $\atWork$h to $\atHome$h when the inhabitants are not present in the building.

The heat-pump thermostat is equipped with sensors to measure the outside temperature and solar irradiation, which are measurements obtained from a location in Belgium~\cite{crawley2001energyplus}.
This work assumes that the heat-pump thermostat is provided with a forecast of the outside temperature and solar irradiation.
However, internal  heat gains caused by the inhabitants and electrical appliances cannot be measured or forecasted and are obtained from~\cite{Dupont}.

\subsection{Learning Agent}

The weights of the auto-encoder network are calculated  at the beginning of each day and  are used to calculate $\batchR$. 
The state information corresponding to the past observations, consists of the previous 10 indoor temperatures and the previous 10 control actions.
The number of hidden neurons of the auto-encoder network is set to 6.
Given the batch $\batchR$, the fitted Q-iteration algorithm constructs a Q-function for the next day (see Algorithm~\ref{FQI}).
This Q-function is then used online by the Boltzmann exploration strategy.
Note that each simulation run begins with an empty batch of tuples and that observations of previous day are daily added to $\batch$.

\subsection{Prescient Method}

The prescient set-back strategy assumes that the model parameters are known and it has prescient knowledge on the outside temperature, solar irradiation, and internal heat gains. 
A detailed description of the prescient method can be found in Appendix C.
The outcome of the prescient set-back strategy is used to evaluate the performance of the learning agent and  can be seen as an absolute lower-bound.

\begin{figure}[t!]
   \begin{center} 
			\includegraphics[width=0.80\columnwidth]{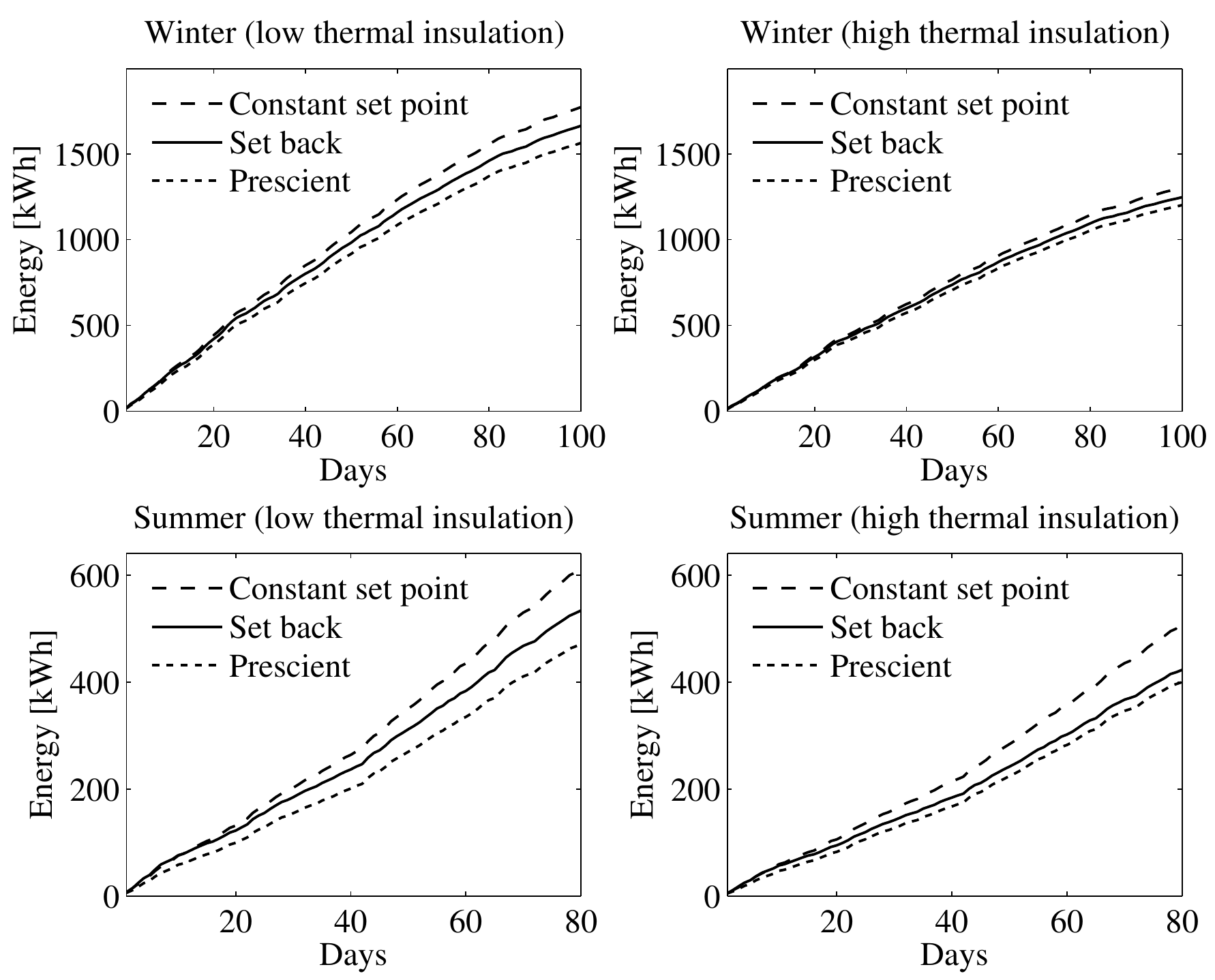}
\caption{Cumulative energy plot of the default constant set-point strategy, our learning agent with set-back strategy and prescient set-back strategy.
}
\label{energies}
 \end{center}
\end{figure}
\begin{figure}[t!]
   \begin{center} 
			\includegraphics[width=0.80\columnwidth]{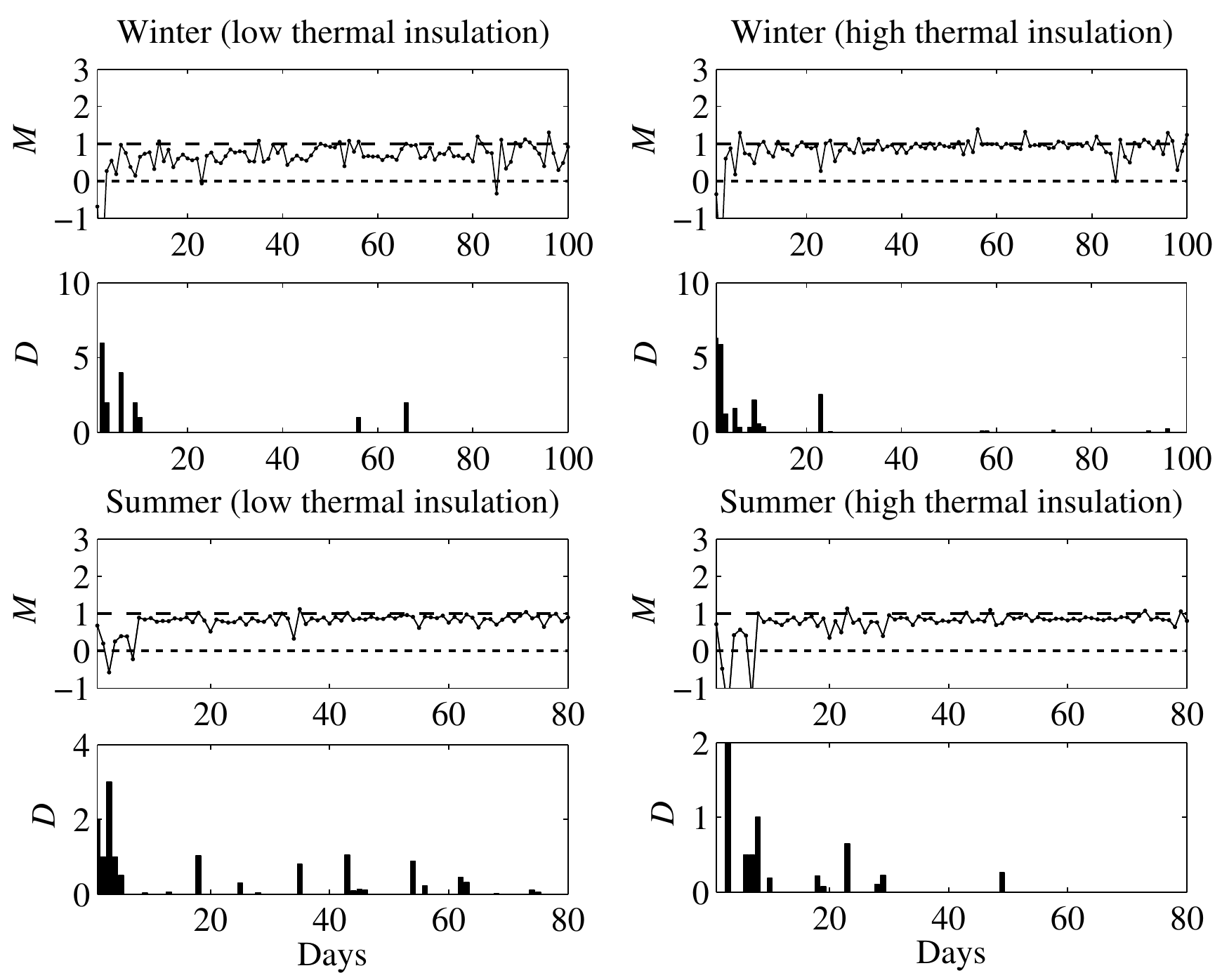}
\caption{The performance metric $M$ and the temperature violations $D$ for the learning agent with set-back strategy.
}
\label{performance}
 \end{center}
\end{figure}

\subsection{Simulation Results}

The experiments compare the energy consumption and temperature violations of the learning agent with a set-back strategy,  conventional constant set-point strategy and  prescient set-back strategy.
Note that the conventional constant temperature set point is the recommended strategy by the U.S. Department of Energy~\cite{DoE}.
In order to examine the adaptability of the learning agent, an identical learning agent is applied to two building types, with a high and low thermal insulation level.
The evaluation is repeated for 100 winter days (heating mode) and 80 summer days (cooling mode).
Figure~\ref{energies} depicts the  cumulative energy consumption of the default controller, prescient controller and learning agent.
As can be seen in Figure~\ref{energies}, the learning agent is able to reduce the total energy consumption compared to the default strategy for both building types.
The simulation results indicate that  the learning agent was able to reduce the energy consumption by $\resultwinter$ during the winter and by $\resultsummer$ during the summer.
It should be noted, however, that the total energy consumption does not give a complete picture, as it does not consider the temperature violations.
Remember that a comfort violation in the heating mode resulted in the activation of the less efficient auxiliary heating element.
However, in the cooling mode no auxiliary cooling is available.
For this reason Figure~\ref{performance} shows the daily performance metric $M_{d}$  and the daily deviation $D_{d}$ between the temperature set point and the indoor temperature at $\atHome$ hour, which is the end of the set-back period.
The daily deviation is calculated as follows:
\begin{equation}
D_{d} = \mathrm{max}(T_{\mathrm{in},17}-\bar{T}_{\mathrm{s},17},0)+\mathrm{max}(\ubar{T}_{\mathrm{s},17}-T_{\mathrm{in},17},0),
\end{equation}
where $T_{\mathrm{in},17}$ is the indoor temperature at 17h, and where  $\ubar{T}_{\mathrm{s},17}$ and $\bar{T}_{\mathrm{s},17}$ are the minimum and maximum temperature set point at 17h.
The daily performance metric $M_{d}$ is calculated as follows:
\begin{equation}
M_{d} = \frac{e_{\mathrm{l}} - e_{\mathrm{d}}}{e_{\mathrm{p}} - {e}_{\mathrm{d}}},
\end{equation}
where $e_{\mathrm{l}}$ denotes the daily energy consumption of the learning agent, $e_{\mathrm{d}}$ denotes the the daily energy consumption of the default strategy and  $e_{\mathrm{p}}$ denotes the daily energy consumption of the prescient controller.
As such, the metric $M$ corresponds to 0 if the learning agent obtains the same performance as the default strategy and corresponds to 1 if the learning agent obtains the same performance as the prescient controller.
These figures show that the comfort violations decrease over the simulation horizon. At the same time the performance metric $M$ increases.

\begin{figure}[t!]
   \begin{center} 
			\includegraphics[width=0.80\columnwidth]{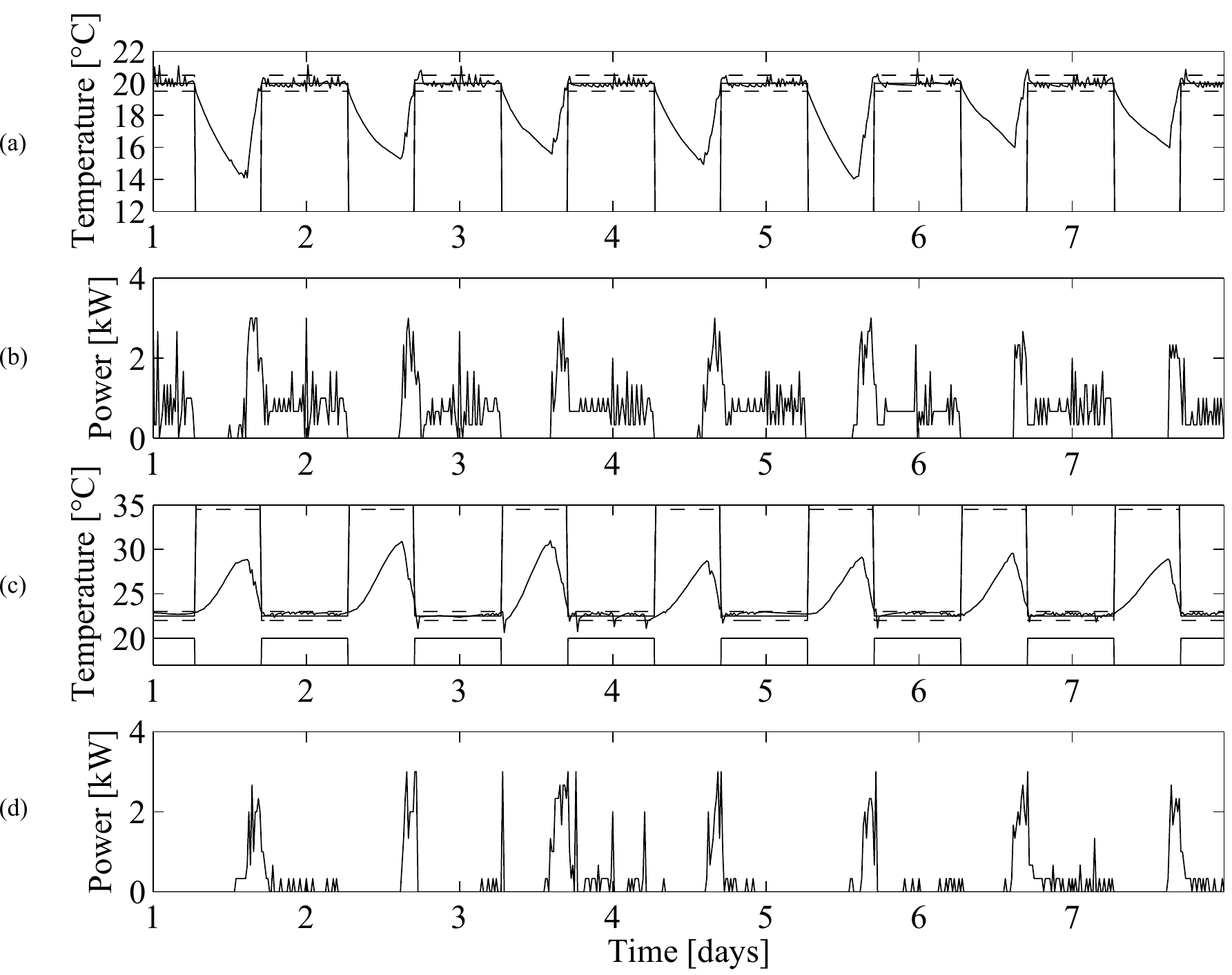}
\caption{Indoor temperature and power consumption of the learning agent with set-back strategy during seven winter (a,b) and summer days (c,d). 
}
\label{temperatures}
 \end{center}
\end{figure}

The  results obtained with a mature controller (batch size of 30 days) are depicted in Figure~\ref{temperatures}.
Figure~\ref{temperatures}.a and Figure ~\ref{temperatures}.b depict the indoor temperature  and power consumption profile during  7 winter days.
Similarly, Figure~\ref{temperatures}.c and Figure~\ref{temperatures}.d depict the indoor temperature and power consumption during 7 summer days.
The simulation results indicate that the proposed learning agent can adapt itself to different building types and outside temperatures.
In addition, the learning agent with  set-back strategy can reduce the energy consumption of a heat pump compared to the conventional constant temperature set-point strategy.

\section{Conclusion and Future Work}
\label{cfw}

This work addressed the challenge of developing a learning agent for a heat pump  with a set-back strategy that saves energy compared to a  constant temperature set-point strategy, which is recommended by the U.S. Department of Energy.
To this end, this paper proposed an approach based on an auto-encoder and a popular model-free batch  reinforcement learning technique, i.e. fitted Q-iteration.
The auto-encoder is used to reduce the dimensionality of the state vector, which  contains past observations of the indoor temperatures and energy consumptions.
The performance of the set-back strategy has been evaluated for heating in the winter and cooling in the summer for two building types with different thermal characteristics.
An equivalent thermal parameter model has been used to simulate the temperature dynamics of the indoor air temperature and the temperature of the building envelope. 
During the winter period, the set-back strategy was able to reduce the energy consumption with $\resultwinter$ compared to the default strategy.
During the summer period, the set-back strategy saved $\resultsummer$ compared to the default strategy.
The results indicated that the proposed learning agent can adapt itself to different building types and weather conditions.
The proposed learning agent obtained these results without making assumptions on the model or its parameters.
As a result, the learning agents can be applied to virtually any building type.

With this work, we intended to show that model-free batch reinforcement learning techniques can provide a valuable alternative to model-based controllers.
In our future work, we plan to focus on including real-time electricity prices in the objective and on implementing the presented approach in a lab environment.

\acknowledgments{Acknowledgments}

Frederik Ruelens has a Ph.D. grant of the Institute for the Promotion of Innovation through Science and Technology in Flanders (IWT-Vlaanderen). KU Leuven is a partner of EnergyVille, Thor park, 3600 Genk, Belgium.


\authorcontributions{Author Contributions}
This paper is part of the doctoral research of Frederik Ruelens, supervised by Ronnie Belmans.
All authors have been involved in the preparation of the manuscript.


\conflictofinterests{Conflicts of Interest}
The authors declare no conflict of interest. 

\subsection*{Appendix A: Thermostat Logic}

Algorithm~\ref{default} illustrates the working of the heat-pump thermostat.
The lower and upper temperature set points are given by $\Tset$ (20 $^\circ C$) and $\Tsetmax$ (22.5 $^\circ C$). 
When the indoor temperature $\Tin$ drops below $\Tset-\Tbufferaux$ (18.5 $^\circ C$ ) the auxiliary heating element is activated in addition to the heat pump until the indoor temperature reaches $\Tset+\Tbuffer$ (20.5 $^\circ C$ ).
The activation of the auxiliary heating is independent of the requested control action and can be seen as an overrule mechanism that guarantees the comfort of the end user.
If the indoor temperature $\Tin$ is between $\Tset+\Tbuffer$ and  $\Tsetmax$ the thermostat controller follows the requested control action.
When the indoor temperature $\Tin$ rises above $\Tsetmax$ the cooling mode is activated until $\Tsetmax-\Tbuffer$ (19.5 $^\circ C$ ). In cooling mode no cooling element is used.

\begin{algorithm}[t]
\caption{ Thermostat of a heat pump with auxiliary heating}
\begin{algorithmic}[1]
\algsetup{linenosize=\tiny}
\STATE Measure $\Tin$  
\STATE Initialize $\Tset, \Tbuffer, \Tbufferaux, \Tsetmax$ 
\STATE \textbf{if}  $\Tin < \Tset - \Tbufferaux$\\
~~ $\uph  = \Pheat +\Paux $  until ${\Tin \geq \Tset +  \Tbuffer}$ 
\STATE \textbf{elseif} $ \Tset - \Tbufferaux\leq \Tin \leq \Tset + \Tbuffer$ \\
~~ $\uph  = \Pheat$\\
\STATE \textbf{elseif} $ \Tset + \Tbuffer < \Tin < \Tsetmax$ \\
~~ $\uph  = u$\\
\STATE \textbf{elseif} $ \Tin \geq \Tsetmax $\\
~~ $\uph  = \Pcool$ until ${\Tin \leq \Tsetmax-   \Tbuffer}$ \\
\STATE \textbf{end if } \\
\end{algorithmic}
\label{default}
\end{algorithm}

\subsection*{Appendix B: Model Equations}

In order to obtain system trajectories of the indoor air temperature, an equivalent thermal parameter model is used to calculate the indoor air temperature $\Tin$ and envelope  temperature $\Tm$ of a residential building~\cite{chassin2008gridlab,Sonderegger}:
\begin{equation}
\begin{matrix}
&\Tindot &=& \frac{1}{\Ca}(\Tm \Hm - \Tin(\Ua+\Hm)+\Qi+\Tout \Ua)\\
&\text{~}\Tmdot &=&\frac{1}{\Cm}(\Hm \left(\Tin-\Tm\right)+\Qm),\text{~~~~~~~~~~~~~~~~~~}
\end{matrix}
\label{eqTCL}
\end{equation}
where $\Hm$ is the  thermal conductance of the building envelope, $\Ua$ is the thermal conductance between air and mass, $\Ca$ is the thermal mass of the air, and  $\Cm$ is the thermal mass of the building and its contents.
The heat added to the interior air mass $\Qi$ is given by a fraction $\alpha$ of the internal heat gains $\Qg$, a fraction $\beta$ of the solar heat gains $\Qs$, and the heat gains generated by the heat pump $\Qhp$.
The heat added to the interior solid mass $\Qm$ is given by the other fractions of $\Qs$ and $\Qg$:
\begin{equation}
\begin{matrix}
&\Qi &=& \alpha \Qg + \beta \Qs + \Qhp\text{~~~~~~~~~~~~~~~~~~~~~~~}\\
&\text{~}\Qm &=& (1-\alpha) \Qg + (1-\beta)\Qs.\text{~~~~~~~~~~~~~~~}
\end{matrix}
\end{equation}
Table~\ref{ETP_parameters} and~\ref{hp_paramters} give the parameters used in the simulations.
The outside temperature $\Tout$ and solar irradiance were obtained from a location in Belgium~\cite{crawley2001energyplus}.
Although more detailed building models exist in the literature~\cite{TRNSYS}, the authors believe that the used model is accurate enough to illustrate the working of the proposed model-free approach and at the same time flexible enough in terms of parameters and computational speed.

\begin{table} [t!]%
\caption{Lumped Parameters}
    \label{ETP_parameters}
\begin{center}
\begin{tabular}{ l  c  c l}
\toprule
 Parameters   & Low  thermal & High thermal&Unit \\
   \midrule
    $\Ua  $    &  1154 &  272      &$ W/{}^\circ C$\\   
    $\Hm $    &  6863 &  6863    & $ W/{}^\circ C$\\   
		$\Ca $   &  2.441 &  2.441     &$MJ/{}^\circ C$\\   
    $\Cm$   &  9.896 &  9.896    & $MJ/{}^\circ C$ \\   
  \bottomrule
\end{tabular}
\end{center}      
\end{table}

\subsection*{Appendix C: Prescient Method}

The optimization problem of the prescient method is formulated as follows:
\begin{equation}
\begin{aligned}
& {\text{minimize}}
& &  \Sigma_{k=1}^{T} \uph \\
& \text{subject to} & &  f(x_{k},u_{k},w) = x_{k+1}. \\
& &   & \uph = T(x_{k},u_{k}),
\end{aligned}
\end{equation}
where the plant model $f$ is defined by~(\ref{eqTCL}) and thermostat $T$ by Algorithm \ref{default}.
In contrast to our model-free approach, the prescient method knows the plant model and future disturbances.
An optimal solution of this optimization problem was found by applying a mixed-integer linear programming solver using Gurobi~\cite{Gurobi} and YALMIP ~\cite{YALMIP}.

\begin{table} [h!]%
\caption{Heat-pump parameters and thermostat settings}
    \label{hp_paramters}
\begin{center}
\begin{tabular}{ l  l  }
\toprule
 Parameters   & Value   \\
   \midrule
    $\Pheat /\ \Pcool$   &  2500 W   \\   
		$\Paux$                 &  3000 W   \\   
    $\Tset $                 &  20 $^\circ C$   \\   
		$\Tsetmax $           &  22.5 $^\circ C$   \\   
    $\Tbuffer$              &     $0.5^\circ C$ \\   
		$\Tbufferaux$         &     $1.5^\circ C$ \\   

  \bottomrule
\end{tabular}
\end{center}      
\end{table}
\bibliographystyle{mdpi}
\makeatletter
\renewcommand\@biblabel[1]{#1. }
\makeatother

\bibliography{references}
\bibliographystyle{mdpi}

\end{document}